\documentclass[EPJ]{svjour}
\journalname{The EPJ Plus}
\usepackage[fleqn]{nccmath}
\usepackage[english]{babel}
\usepackage{amsmath}
\usepackage{amssymb}
\usepackage{upgreek}
\usepackage[T1]{fontenc}
\usepackage{color}
\allowdisplaybreaks
\usepackage{mathtools}
\usepackage{eucal}
\usepackage{graphicx}
\usepackage{float}
\usepackage{url}
\usepackage{epsfig}
\usepackage{bm}
\usepackage[utf8x,latin1]{inputenc} 
\usepackage{emerald}
\usepackage{slantsc}
\usepackage{array}
\usepackage[sort&compress,numbers]{natbib}

\def\be{\begin{equation}}
\def\ee{\end{equation}}
\def\ba{\begin{eqnarray}}
\def\eqna{\end{eqnarray}}
\def\bsu{\begin{subequations}}
\def\esu{\end{subequations}}

\def\pd{\partial}
\def\le{\left}
\def\ri{\right}                        

\begin{document}
\onecolumn
\title{The Orbital Pericenter Precession in the 2PN Approximation}
\author{Sergei M. Kopeikin 
\thanks{https://orcid.org/0000-0002-4866-1532}
	}                     
\institute{Department of Physics \& Astronomy, University of Missouri, 322 Physics Bldg., Columbia, Missouri 65211, USA\\
\email{kopeikins@missouri.edu}}
\date{Received:               / Revised version: }

\abstract{
Recent article {\it Revisiting the 2PN Pericenter Precession in View of Possible Future Measurements} published by Iorio (Universe 2020) argues that calculations of the secular 2PN precession of the orbital pericenter of a binary system accomplished by Damour \& Sch\"afer (Nuovo Cim. B 1988)\nocite{damour_schafer_1988NCimB}
and by Kopeikin \& Potapov (Astron. Rep 1994)\nocite{kopeikin_potapov_1994ARep} and  with different mathematical techniques are inconsistent, differ from each other and don't agree with the result obtained by Iorio. The purpose of this communication is to demonstrate that the article by Iorio (Universe 2020)\nocite{iorio_2020} is erroneous, suffers from misconceptions and cannot be trusted.     
} 
\maketitle
\tableofcontents
\section{Introduction}\label{sec1}
Relativistic effect of a secular motion of pericenter of an elliptical orbit of two-body system is one of the fundamental tests of general relativity \citep{willLRR}. It was calculated for the orbit of Mercury in the first post-Newtonian (1PN) approximation by \citet{Einstein_1915SPAW} who considered Mercury as a test body moving in a spherically-symmetric gravitational field of the Sun. Einstein's calculation were generalized to the system of two bodies of comparable masses by \citet{Robertson_1938AnMat} by solving the Einstein-Infeld-Hoffmann (EIH) equations of motion \citep{EIH_1938AnMat}. 

Discovery of a binary pulsar PSR 1913+16 \citep{hulse-taylor_1975} and rapid development of pulsar timing technique revealed a fascinating opportunity to test general relativity in higher-order PN approximations of general relativity in strong-field regime \citep{damour_taylor_PRD1992}. Equations of motion of two-body system in the 2PN and 2.5PN approximations had been derived by applying various mathematical methods by \citet{Schafer_1982PThPh,Damour_1983grr,k85} who also demonstrated that the resulting equations of motion and corresponding Lagrangians are in complete agreement \citep{Damour_1989MG5}. 

\citet{damour_schafer_1988NCimB} integrated the equations of motion of two-body problem by the Hamilton-Jacobi method in the ADM coordinates and derived a secular advance of the orbital pericenter extending the Einstein-Robertson formula \citep{Robertson_1938AnMat} to the 2PN approximation. Later on,
\citet{kopeikin_potapov_1994ARep} integrated equations of motion of a binary system in harmonic coordinates in the 2PN approximation by the method of osculating elements which is a standard technique of dynamic astronomy and celestial mechanics \citep{kovalevsky_book}. They have demonstrated that the osculating elements technique yields exactly the same result as that obtained earlier by \citet{damour_schafer_1988NCimB} after taking into account the post-Newtonian transformation between the ADM and harmonic coordinates and the constants of integration adopted in \citep{kopeikin_potapov_1994ARep} and \citep{damour_schafer_1988NCimB}.           

Recent article by \citet{iorio_2020} attempted to derive the secular shift of Mercury's perihelion in the 2PN approximation in view of its possible measurement with BepiColombo spacecraft mission \citep{iess_2018cosp}. \citet{iorio_2020} has considered motion of a test particle (Mercury) in spherically-symmetric field of the Sun and integrated the geodesic equations of motion of the particle in the harmonic coordinates by the method of osculating elements but with a different choice of the constants of integration referred to an initial epoch $t_0$. \citet{iorio_2020} has tried but completely failed to establish a correspondence of his result with \citep{damour_schafer_1988NCimB} and with \citep{kopeikin_potapov_1994ARep} and claimed that the result of \citet{kopeikin_potapov_1994ARep} is erroneous in spite that it exactly coincides with the result by \citep{damour_schafer_1988NCimB}. \citet{iorio_2020} also failed to see that two expressions for the 2PN pericenter advance given in paper by \citet{damour_schafer_1988NCimB}, are identical and claimed that they are self-contradictory.  

The goal of our communication is to show that paper by \citet{iorio_2020} is full of misconceptions, clogged with mathematical flaws and is wrong. In sections \ref{sec2} and \ref{sec3} we briefly outline the results of calculations of the 2PN secular pericenter advance obtained by \citet{damour_schafer_1988NCimB} and \citet{kopeikin_potapov_1994ARep} and prove their identity in section \ref{sec4}. Section \ref{sec5} explains mathematical mistakes of the paper \citep{iorio_2020}. Section \ref{sec6} summarizes our discussion of the 2PN pericenter precession and concludes that the paper by \citet{iorio_2020} cannot be trusted. 

\section{The 2PN secular pericenter advance derived by the Hamilton-Jacobi method}\label{sec2}

The 2PN orbital parametrization of the relative orbit of two-body system chosen by \citet{damour_schafer_1988NCimB} is 
\ba\label{1}
2\pi\frac{t-t_0}{P}&=&u-e_t\sin u+\frac{f}{c^4}\sin v+\frac{g}{c^4}(v-u)\;,\\\label{2}
r&=&a_R\le(1-e_R\cos u\ri)\;,\\\label{3}
2\pi\frac{\varphi-\varphi_0}{\Phi}&=&v+\frac{F}{c^4}\sin 2v+\frac{G}{c^4}\sin 3v\;,
\eqna
where
\ba\label{4}
v&=&2\arctan\le[\le(\frac{1+e_\varphi}{1-e_\varphi}\ri)^{1/2}\tan\frac{u}2\ri]\;.
\eqna
Here $t$ is time, $t_0$ is the initial epoch, $\varphi$ is the phase angle in the orbital plane counted from $\varphi_0$ - the initial phase, $r$ is the radial distance between the two bodies, $u$ and $v$ are the eccentric and true anomalies, $a_R$ and $e_R$ are constant semi-major axis and eccentricity of the orbit, $e_t$ and $e_\varphi$ are two other constant eccentricities, $P$ and $\Phi$ are the constant period and angle of the orbital return to the pericenter, and $f,g,F,G$ are some constants which are not important for the goal of the present paper.

The constant parameters $P$ and $\Phi$ entering the orbital parametrization \eqref{1}--\eqref{4} can be found with the the Hamilton-Jacobi method. This method utilizes the action $S$ written in the form \citep{Landau_machanics} 
\ba\label{5}
S&=&-{\cal E}t+h\varphi+i_r({\cal E},h)\;,
\eqna  
where ${\cal E}$ and $h$ are the orbital energy and angular momentum (per unit mass) of the relative motion in two-body system, and $i_r({\cal E},h)$ is a radial action integral \citep[eq. 3.10]{damour_schafer_1988NCimB}
\ba\label{6}
i_r({\cal E},h)&=&
-h+\frac1{c^2}\frac3{h}+\frac1{c^4}\le[\left(\frac{35}4-\frac52\nu\ri)\frac1{h^3}+\le(\frac{15}2-3\nu\ri)\frac{{\cal E}}{h}\ri]\\\nonumber
&&+\frac1{\sqrt{-2{\cal E}}}\left[1+\frac1{c^2}\le(\frac{15}4-\frac14\nu\ri){\cal E}+\frac1{c^4}\le(\frac{35}{32}+\frac{15}{16}\nu+\frac3{32}\nu^2\ri){\cal E}^2\ri]\;,
\eqna
and $\nu\equiv m_1m_2/m^2$ with $m_1$ and $m_2$ being masses of the bodies in two-body system while $m\equiv m_1+m_2$ is the total mass.

According to the Hamilton-Jacobi method parameters $P$ and $\Phi$ are defined by equations
\ba\label{8}
\frac{P}{2\pi Gm}&=&+\frac{\pd}{\pd {\cal E}}i_r({\cal E},h)\;,\\
\label{7}
\frac{\Phi}{2\pi}&=&-\frac{\pd}{\pd h}i_r({\cal E},h)\;.
\eqna
Taking the partial derivatives yield
\ba\label{7a}
\frac{1}{nGm}&=&\le(-2{\cal E}\ri)^{-3/2}\le[1-\frac14(15-\nu)\frac{{\cal E}}{c^2}-\frac{3}{32}(35+30\nu+3\nu^2)\frac{{\cal E}}{c^4}+\frac32(5-2\nu)\frac{(-2{\cal E})^{3/2}}{c^4h}\ri]\;,\\
\label{10}
k&=&\frac{3}{c^2h^2}\le[1+\le(\frac52-\nu\ri)\frac{{\cal E}}{c^2}+\le(\frac{35}{4}-\frac52\nu\ri)\frac{1}{c^2h^2}\ri]\;.
\eqna
where 
\ba\label{8a}
n&\equiv &\frac{2\pi}{P}\;,
\eqna
is the orbital frequency of the orbital radial oscillations, and
\ba
\label{9}
k&\equiv&\frac{\Delta\Phi}{2\pi}\equiv\frac{\Phi-2\pi}{2\pi}\;,
\eqna
is the dimensional parameter measuring the fractional pericenter advance per orbit.  

Formula \eqref{10} is the theoretical prediction for the 2PN secular pericenter advance expressed in terms of the two constant parameters - the reduced energy ${\cal E}$ and the angular momentum $h$ - which are the first integrals of motion. However, it is not convenient for interpretation of experimental data since the orbital energy and angular momentum are not directly observable. Therefore, \citet{damour_schafer_1988NCimB} expressed $k$ in terms of the observables which are the orbital frequency $n$, the eccentricity of the orbit of the body with mass $m_1$ with respect to the center of mass of the two-body system $e_{\rm T}$ (see \citep[page 272]{DD_II_1986} for more detail), and masses of the bodies -- $x_1=m_1/m$, $x_2=m_2/m$ -- normalized to the total mass $m=m_1+m_2$. In terms of these parameters formula \eqref{10} takes on the following form \citep[eq. 5.18]{damour_schafer_1988NCimB}
\ba\label{11}
k&=&\frac{3\le(Gmn\ri)^{2/3}}{c^2(1-e_{\rm T}^2)}\le[1+\frac{\le(Gmn\ri)^{2/3}}{c^2(1-e_{\rm T}^2)}\le(\frac{39}4x^2_1+\frac{27}4x^2_2+15x_1x_2\ri)-\frac{\le(Gmn\ri)^{2/3}}{c^2}\le(\frac{13}4x^2_1+\frac{1}4x^2_2+\frac{13}{3}x_1x_2\ri)\ri]\;.
\eqna
It is worth observing that the product $x_1x_2\equiv\nu$ and $x_1+x_2\equiv 1$.

The conversion from \eqref{10} to \eqref{11} is done with the help of post-Newtonian transformations between various parameters of the two-body system, derived in \citep{DD_I_1985} 
\ba\label{12}
{\cal E}&=&-\frac{Gm}{2a_R}\le[1-\frac14\le(7-\nu\ri)\frac{Gm}{c^2a_R}\ri]\;,
\\
\label{13}
n&=&\le(\frac{Gm}{a_R^3}\ri)^{1/2}\le[1+\frac12\le(\nu-9\ri)\frac{Gm}{c^2a_R}\ri]\;,
\\
\label{14}
h&=&\le[\frac{a_R(1-e_R^2)}{Gm}\ri]^{1/2}\le[1-\frac{Gm}{c^2a_R}\le(1-\frac12\nu\ri)+\frac{Gm}{c^2a_R(1-e^2_R)}\le(3-\frac12\nu\ri)\ri]\;.
\eqna

It should be noticed that the orbital parametrization of the two-body problem chosen by \citet{damour_schafer_1988NCimB} in the 2PN approximation is a direct extension of the 1PN parametrization worked out by \citet{DD_I_1985,DD_II_1986}. It has five different eccentricities denoted as $e_t$, $e_{\rm T}$, $e_\varphi$, $e_r$ and $e_R$. Among them, the eccentricity $e_t$ appears in the Kepler equation of time \eqref{1}, the eccentricity $e_R$ characterizes the oblateness of the relative orbit in equation \eqref{2}, and the eccentricity $e_\varphi$ relates the angular variables (anomalies) by means of equation \eqref{3} \footnote{The variable $\varphi\equiv\theta$ and eccentricity $e_\varphi\equiv e_\theta$ in the notations of the paper \citep{DD_I_1985}.}. 
Two remaining eccentricities are intimately related to the orbit of the first body with respect to the center of mass of the two-body problem. The eccentricity
\ba\label{15}
e_r&=&e_R\le[1+\frac{Gm}{2c^2a_R}\le(x_1x_2-x^2_1\ri)\ri]\;,
\eqna
where $x_1$ and $x_2$ have been defined above.
The eccentricity $e_{\rm T}$, which appears in the 2PN secular pericenter advance formula \eqref{11}, is a linear combination of the other three eccentricities \citep{DD_II_1986}
\ba\label{16}
e_{\rm T}&=&e_t(1+\delta)+e_\varphi-e_r\;,
\eqna
where 
\ba\label{17}
\delta&=&\frac{Gm}{c^2a_R}\le(x_1x_2+2x^2_2\ri)\;,
\eqna
is the parameter characterizing the amplitude of the post-Newtonian periodic variations between the proper time ${\rm T}$ of the body with mass $m_1$ and the coordinate time $t$.    

\section{The 2PN secular pericenter advance derived by the method of osculating elements}\label{sec3}

Integration of relativistic equations of motion of two-body problem in the 2PN approximation has been performed by \citet{kopeikin_potapov_1994ARep} in 1994. We have undertaken this study in order to check the results of the paper \citep{damour_schafer_1988NCimB} by making use of independent mathematical technique which is often employed by the students of classic celestial mechanics. The osculating elements are not the optimal instrument for the integration of relativistic equations of motion because they lead to the appearance of terms with multiple orbital frequencies which are, in fact, unobservable and can be eliminated by using transformations of the constants of integration. Nonetheless, it is straightforward to use the technique of the osculating elements to get an independent consistency check of the results obtained by other methods.

The method of osculating elements postulates that the perturbed orbit of two-body system preserves the same form as in classic Newtonian theory but with all orbital parameters depending on time \citep{kovalevsky_book}
\ba\label{18}
M&=&E-e\sin E\;,\\\label{19}
r&=&a\le(1-e\cos E\ri)\;,\\\label{20}
\theta&=&f+\omega\;,
\eqna
where $a$ is the semi-major axis, $e$ is eccentricity, the angle $\theta$ is called the argument of latitude, and the angles $f$, $E$ and $M$ are true, eccentric and mean anomalies, respectively \footnote{Notation $E$ for the eccentric anomaly should not be confused with that for the reduced orbital energy ${\cal E}$ per unit mass.}. The argument of longitude $\theta$ is reckoned from the ascending node $\Omega$ of the orbit, while the true anomaly $f$ is reckoned from the pericenter, $\omega$, of the orbit. The true and eccentric anomalies are interrelated by equation
\ba\label{21}
f&=&2\arctan\le[\le(\frac{1+e}{1-e}\ri)^{1/2}\tan\frac{E}2\ri]\;,
\eqna
where the eccentricity $e$ is the same as that in \eqref{19}, \eqref{20}.
The mean anomaly $M$ depends on time as follows,
\ba\label{22}
M&=&M_0+{\sf n}\le(t-t_0\ri)\;,
\eqna
where $t_0$ is the initial epoch, $M_0$ is the mean anomaly at the epoch, and 
\ba\label{23}
{\sf n}&=&\le(\frac{Gm}{a^3}\ri)^{1/2}  \;,
\eqna
is the frequency of the radial motion of the osculating orbit. 

In case of a two-body system the orbital elements $a$, $e$, $\omega$ and frequency ${\sf n}$ are constant in the Newtonian theory but they become functions of time in the post-Newtonian approximations of general relativity \citep{vab,sof89}. Their exact form in the 1PN approximation can be found, for example, in paper by \citet{klioner_kopeikin_1994ApJ}. Solution for the osculating elements in the 2PN approximation of general relativity was found in \citep{kopeikin_phd,kopeikin_potapov_1994ARep}. In particular, the orbital pericenter $\omega=\omega_0+kf+\mbox{periodic terms}$ has a secular component (precession) 
\ba\label{23a}
k&\equiv&\le<\frac{d\omega}{df}\ri>=\frac1{2\pi}\int_0^{2\pi}\frac{d\omega}{df}df\;,
\eqna
that is the average value of the pericenter advance per one orbital revolution \footnote{Notice that the angular variable $f$ is measured with respect to the perturbed (precessing) position of the pericenter as defined by the method of osculating elements \citep{kovalevsky_book} -- for more detail see \citep[Eq. 11]{kopeikin_potapov_1994ARep}.}.
We have shown \citep[eq. 5.2]{kopeikin_potapov_1994ARep} that
\ba\label{24}
k&=& \frac{3Gm}{c^2\mathtt{a}(1-\mathtt{e}^2)}\le[1+ \frac{Gm}{c^2\mathtt{a}(1-\mathtt{e}^2)}\le(\frac34+\frac32\nu\ri)-\frac{Gm}{c^2\mathtt{a}}\le(\frac14+\frac52\nu\ri) \ri]\;, 
\eqna
where $\mathtt{a}$ and $\mathtt{e}$ are constants of integration which are the mean values of the perturbed orbital elements, $a=a(f)$ and $e=e(f)$, over one orbital period with respect to the (perturbed) true anomaly $f$,
\ba\label{25}
\mathtt{a}\equiv\frac1{2\pi}\int_0^{2\pi}\frac{da}{df}df\qquad\;, \qquad\mathtt{e}\equiv\frac1{2\pi}\int_0^{2\pi}\frac{de}{df}df\;.
\eqna

Sometimes the constants of integration are chosen differently as the values $a_0\equiv a(f_0)$ and $e_0\equiv e(f_0)$ of the osculating elements taken at the initial instant of time $t_0$ corresponding to the initial value of the true anomaly $f_0$. The constant mean values of the orbital elements $\mathtt{a}$ and $\mathtt{e}$ differ from the constant initial values of $a_0$ and $e_0$ by the post-Newtonian terms \citep{klioner_kopeikin_1994ApJ}
\ba\label{ki7p}
a_0&=&\mathtt{a}+da_0\qquad\;,\qquad e_0=\mathtt{e}+de_0\;,
\eqna
where 
\ba\label{okw4c}
da_0&=&\frac{Gm\mathtt{e}}{c^2(1-\mathtt{e}^2)^2}\le\{\le[-14+6\nu+\mathtt{e}^2\le(-6+\frac{31}4\nu\ri)\ri]\cos f_0+ \mathtt{e}(-5+4\nu)\cos 2f_0+\frac14 \mathtt{e}^2\nu \cos 3f_0\ri\} \;,\\\label{kjev3}
de_0&=&\frac{Gm}{c^2\mathtt{a}(1-\mathtt{e}^2)}\le\{ \le[-3+\nu+\mathtt{e}^2\le(-7+\frac{47}8\nu\ri)\ri]\cos f_0+ \frac12\mathtt{e}(-5+4\nu)\cos 2f_0+\frac18 \mathtt{e}^2\nu \cos 3f_0\ri\}\;.
\eqna 

Here is the issue which must be clearly articulated and understood in order to avoid confusions and misinterpretations of the iterative solution of the Gauss equations of the osculating elements in relativistic theory. More specifically, we emphasize that the mean values of the osculating elements, $\mathtt{a}$ and $\mathtt{e}$, are constants of integration which don't depend on the initial phase $f_0$ {\it at all}. It means that they are intimately related to the first integrals of motion, ${\cal E}$ and $h$, in the sense that there is one-to-one mapping \footnote{One-to-one mapping (or bijection) means that each element of one set is paired with exactly one element of the other set, and each element of the other set is paired with exactly one element of the first set \url{https://en.wikipedia.org/wiki/Bijection}.} between the two sets of constants: ${\cal E}={\cal E}(\mathtt{a},\mathtt{e})$ and $h=h(\mathtt{a},\mathtt{e})$ -- see equations \eqref{12a}, \eqref{14a} below. On the other hand, the constants of integration $a_0=a(f_0)$ and $e_0=e(f_0)$ are functions of the initial phase $f_0$ as shown in \eqref{ki7p}--\eqref{kjev3}. Therefore, they don't remain constant if one shifts the initial phase $f_0$ to another numerical value, say $f^*_0$, without changing the geometric shape of the particle's orbit. Thus, there is no one-to-one mapping between the two sets $\{{\cal E}, h\}$ and $\{a_0,e_0\}$. In order to relate the integrals of motion ${\cal E}$ and $h$ with the initial data set of the osculating elements $\{a_0,e_0\}$ one has to include to this set the initial phase $f_0$ as an additional parameter. It makes the integrals of motion functions of three parameters: ${\cal E}={\cal E}(a_0,e_0,f_0)$ and $h=h(a_0,e_0,f_0)$. If one denotes ${\cal E}_0=:{\cal E}(a_0,e_0)$ and $h_0\equiv h(a_0,e_0)$, then
the post-Newtonian expansion of the orbital elements yields, 
\ba\label{int123}
{\cal E}&=&{\cal E}_0-da_0\frac{\pd{\cal E}}{\pd\mathtt{a}}-de_0\frac{\pd{\cal E}}{\pd\mathtt{e}}+\ldots\;,\\
\label{int124}
h&=&h_0-da_0\frac{\pd h}{\pd\mathtt{a}}-de_0\frac{\pd h}{\pd\mathtt{e}}+\ldots\;,
\eqna
where $da_0$ and $de_0$ are given in \eqref{okw4c}, \eqref{kjev3} respecively, and the lower dots symbol $\ldots$ denotes the residual terms of the expansion. The terms ${\cal E}$ and $h$ in the left hand side of \eqref{int123}, \eqref{int124} are first integrals of motion which are true constants of integration that do not depend on the choice of the initial position of particle (the initial phase $f_0$) on its orbit. At the same time, the "unperturbed" initial values of the integrals of motion, ${\cal E}_0$ and $h_0$ are constants only approximately for they change their values if one shifts the initial phase in order to compensate the change of the terms with the partial derivatives $\pd{\cal E}_0/\pd\mathtt{a}$, $\pd h/\pd\mathtt{a}$, etc. This issue was not properly understood and is completely misinterpreted by \citet{iorio_2020}. We discuss Iorio's mistreatment of the problem of the constants of integration in section \ref{sec5} in more detail.     

\section{Proof of the identity of Damour-Sch{\"a}fer [1] and Kopeikin-Potapov [2] results for the 2PN pericenter advance.}\label{sec4} 
Definition \eqref{23a} for the secular 2PN advance of the orbital pericenter of two-body system given by \citet{kopeikin_potapov_1994ARep} in terms of the osculating elements has the same meaning as definition \eqref{9} taken from the paper by \citet{damour_schafer_1988NCimB} and given in terms of the Hamilton-Jacobi elements. Hence, the two equations, \eqref{10} and {\eqref{24}, for the secular advance of the pericenter must coincide. In order to see that they do coincide we notice that  
the orbital elements $a_R$ and $e_R$ in paper by \citet{damour_schafer_1988NCimB} are related to those in the paper by \citet{kopeikin_potapov_1994ARep} by the following equations \citep[eqs. 28 and 29]{klioner_kopeikin_1994ApJ}
\ba\label{26}
a_R&=&\mathtt{a}+\frac{Gm}{c^2(1-\mathtt{e}^2)^2}\le[3-\nu+\mathtt{e}^2(13-7\nu)-\mathtt{e}^4(1+2\nu)\ri]\;,
\\\label{27}
e_R&=&\mathtt{e}+\frac{Gm\mathtt{e}}{2c^2\mathtt{a}(1-\mathtt{e}^2)}\le[17-6\nu-2\mathtt{e}^2(1+2\nu)\ri]\;.
\eqna
Making use of these transformations we can express the orbital energy \eqref{12} and angular momentum \eqref{14} in terms of the constants of integration of the osculating elements. With sufficient accuracy we have
\ba\label{12a}
{\cal E}&=&-\frac{Gm}{2\mathtt{a}}\le\{1-\frac{Gm}{c^2a(1-\mathtt{e}^2)}\le[19-5\nu+2\mathtt{e}^2(19-13\nu)+3\mathtt{e}^4(1-3\nu)  \ri]   \ri\}\;,
\\
\label{14a}
h&=&\le[\frac{\mathtt{a}(1-\mathtt{e}^2)}{Gm}\ri]^{1/2}\le[1-\frac{Gm}{2c^2\mathtt{a}}\le(1-3\nu\ri)+\frac{2Gm}{c^2\mathtt{a}(1-\mathtt{e}^2)}\le(2-\nu\ri)\ri]\;.
\eqna
Replacing ${\cal E}$ and $h$ in \eqref{10} with equations \eqref{12a} and \eqref{14a} one gets exactly equation \eqref{24} that establishes their identity. 

What remains is to show that equation \eqref{11} for $k$ derived by \citet{damour_schafer_1988NCimB} also coincides with our equation \eqref{24}. To this end we, first, express equation \eqref{11} in terms of the parameters $\nu$ and $x_2$ instead of $x_1$ and $x_2$ by observing that $x_1x_2=\nu$ and $x^2_1=1-2\nu-x^2_2$. We get,
\ba\label{oln5}
k&=&\frac{3\le(Gmn\ri)^{2/3}}{c^2(1-e_{\rm T}^2)}\le[1+\frac{\le(Gmn\ri)^{2/3}}{c^2(1-e_{\rm T}^2)}\le(\frac{39}4-\frac{9}{2}\nu\ri)-\frac{\le(Gmn\ri)^{2/3}}{c^2}\le(\frac{13}4-\frac{13}{6}\nu\ri)-\frac{3\le(Gmn\ri)^{2/3}}{c^2(1-e_{\rm T}^2)}e_{\rm T}^2x^2_2\ri]\;.
\eqna
Second, we express equations \eqref{13} and \eqref{16} in terms of the integration constants $\mathtt{a}$ and $\mathtt{e}$ of the osculating elements. It can be done again by making use of transformation formulas for the constants of integrations in different parametrization of two-body problem from the paper \cite{klioner_kopeikin_1994ApJ}. More specifically, we have the orbital frequency \eqref{13} transformed with the help of \eqref{26} to
\ba\label{13a}
n&=&\le(\frac{Gm}{\mathtt{a}^3}\ri)^{1/2}\le\{1-\frac{Gm}{2c^2\mathtt{a}(1-\mathtt{e}^2)^2}\le[18-4\nu+\mathtt{e}^2(21-19\nu)+\mathtt{e}^4(6-7\nu)\ri]\ri\}\;.
\eqna
As for the eccentricity $e_{\rm T}$ defined in \eqref{16}, we need the following expressions \citep{klioner_kopeikin_1994ApJ}
\ba\label{za28}
e_t&=&\mathtt{e}-\frac{Gm\mathtt{e}}{2c^2\mathtt{a}(1-\mathtt{e}^2)}\le[9-3\nu+\mathtt{e}^2(6-7\nu)\ri]\;,\\
\label{hg25}
e_\varphi&=&\mathtt{e}+\frac{Gm\mathtt{e}}{2c^2\mathtt{a}(1-\mathtt{e}^2)}\le[-17+5\nu+\mathtt{e}^2(2+5\nu)\ri]\;.
\eqna
Substituting these expressions along with \eqref{15} to definition \eqref{16}, we get
\ba
\label{16a}
e_{\rm T}&=&\mathtt{e}\le\{1-\frac{Gm}{2c^2\mathtt{a}(1-\mathtt{e}^2)}\le[10-3\nu+\mathtt{e}^2(5-7\nu)\ri]+\frac{3Gm}{2c^2\mathtt{a}}x^2_2\ri\}\;.
\eqna
Inserting equations \eqref{13a}, \eqref{16a} into \eqref{oln5} and reducing similar terms one can find out that the outcome exactly coincides with expression \eqref{24} for $k$. This proves that equation \eqref{11} derived by \citet{damour_schafer_1988NCimB} for the secular advance of pericenter of two-body system by the Hamilton-Jacobi method, is identical to equation \eqref{24} obtained by \citep{kopeikin_potapov_1994ARep} with the help of the method of osculating elements \footnote{Our study of the pericenter advance in the system of two bodies of comparable masses is fully consistent in the limit of a test particle with an independent post-Newtonian analysis of the pericenter advance conducted by \citet{will_2019CQG}. }.

\section{Misconceptions and mathematical errors in the article by Iorio [3]}\label{sec5}

Article by \citet{iorio_2020} is an attempt to repeat calculations of the 2PN pericenter advance which has been already done by \citet{damour_schafer_1988NCimB} and \citet{kopeikin_potapov_1994ARep}. \citet{iorio_2020} was not able to reproduce the result obtained in \citep{damour_schafer_1988NCimB,kopeikin_potapov_1994ARep} due to misconceptions and mathematical errors which we discuss in this section.
\begin{itemize}
\item[(1)]  Formula (57) in \citep{iorio_2020} is written as follows
\ba\label{ior1}
\frac{\Delta\omega^{\rm 2PN}_{\rm tot}}{2\pi}&=& \frac{3\le(Gmn\ri)^{2/3}}{c^2(1-e_t^2)}\le[1+\frac{\le(Gmn\ri)^{2/3}}{c^2(1-e_t^2)}\le(\frac{39}4x^2_1+\frac{27}4x^2_2+15x_1x_2\ri)-\frac{\le(Gmn\ri)^{2/3}}{c^2}\le(\frac{13}4x^2_1+\frac{1}4x^2_2+\frac{13}{3}x_1x_2\ri)\ri]
\eqna
\citet{iorio_2020} mistakenly assumes that this formula is exactly the same one as given by \citep[eq. 5.18]{damour_schafer_1988NCimB} -- see \eqref{11}. However, the eccentricity $e_t$ used by \citet[eq. 68]{iorio_2020} in \eqref{ior1} is the same as given in \eqref{za28} but this is not the same as $e_{\rm T}$ given in \eqref{16a} which was used in paper by \citet{damour_schafer_1988NCimB}. Apparently, $e_t\neq e_{\rm T}$. Due to the misinterpretation of the eccentricities, $e_{\rm T}$ and $e_t$, \citet{iorio_2020} came to a wrong conclusion that equations \eqref{10} and \eqref{11} are in disagreement \footnote{See the sentence between equations (67) and (68) in \citep{iorio_2020}.} but they are not as we have demonstrated above in section \ref{sec4}. Formula \eqref{ior1} is wrong and the mathematical operations performed by \citet{iorio_2020} after his equations (57) in order to get his equation (70)  make no sense. Iorio's equation (70) in \citep{iorio_2020} is absolutely wrong.    

\item[(2)] On the other hand, equation (53) in \citet{iorio_2020} 
\ba\label{ior2}
\frac{\Delta\omega^{\rm 2PN}_{\rm tot}}{2\pi}&=&\frac{3G^2m^2}{4c^4 a_0^2}\frac{2+e_0^2-32e_0\cos f_0}{(1-e_0^2)^2}\;,
\eqna
is correct in contrast to what \citet{iorio_2020} states. Indeed, equation \eqref{ior2} can be directly reproduced from our equation \eqref{24} which exactly coincides with the equations \eqref{10} and/or \eqref{11} of the paper \citep{damour_schafer_1988NCimB} after expressing it in terms of the constants of integration $\{a_0, e_0, f_0\}$ referred to the initial epoch $t_0$ with the help of \eqref{ki7p}--\eqref{kjev3} and taking the test-mass limit: $m_1\rightarrow 0$ and $m_2\rightarrow m$. 

\item[(3)] Equation \eqref{ior2} should be interpreted with care for it cannot be treated separately from the contribution of the 1PN term $\Delta\omega^{\rm 1PN}_{\rm tot}$. More precisely, the overall secular precession of orbit per one orbital revolution is given by 
\ba\label{ior3}
k&=&\frac{\Delta\omega^{\rm 1PN}_{\rm tot}}{2\pi}+\frac{\Delta\omega^{\rm 2PN}_{\rm tot}}{2\pi}\;,
\eqna
where 
\ba\label{ior4}
\frac{\Delta\omega^{\rm 1PN}_{\rm tot}}{2\pi}&=&\frac{3Gm}{c^2a_0(1-e_0^2)}\;,
\eqna
and $\Delta\omega^{\rm 2PN}_{\rm tot}$ is given by \eqref{ior2}. \citet{iorio_2020} misinterprets the term $\Delta\omega^{\rm 1PN}_{\rm tot}$ as a constant being independent of the initial conditions as it does not depend explicitly on the initial phase $f_0$. He notes, however, that the 2PN term $\Delta\omega^{\rm 2PN}_{\rm tot}$ depends on the initial orbital phase $f_0$ and changes if the initial phase is shifted to another epoch. From this observation, \citet{iorio_2020} draws a wrong conclusion that the 2PN secular pericenter precession can have different values depending on the choice of the initial conditions. This conclusion is based on the misinterpretation of the main term $\Delta\omega^{\rm 1PN}_{\rm tot}$ as constant. It is not constant since it depends on the initial values of $a_0$ and $e_0$ which are functions of the initial orbital phase $f_0$ in accordance with \eqref{ki7p}--\eqref{kjev3}. Hence, if one shifts the initial phase $f_0$ to another value both $\Delta\omega^{\rm 1PN}_{\rm tot}$ and $\Delta\omega^{\rm 2PN}_{\rm tot}$ will change synchronously in such a way that the total precession $k$ of the orbit remains constant. It agrees with the fact that according to the Hamilton-Jacobi method of integration the total precession $k$ is the integral of radial motion \eqref{10} depending only on constant values of energy ${\cal E}$ and orbital angular momentum $h$ which don't depend on the choice of the initial conditions at all. 
\item[(4)] The preceding item helps us to point out to a major mathematical error committed by \citep{iorio_2020} in calculations of the secular pericenter precession. Indeed, let us look at the result of \citet{iorio_2020} for the precession which is given in the limit of a test mass orbiting a central body with mass $m$. It represents a linear combination of the 1PN and 2PN precessional terms with the 2PN precession consisting of the sum of the "direct" and "indirect" precessional formulas given by equations (8) and (23) in \citep{iorio_2020}. The net result of the pericenter precession is
\ba\label{ior5}  
k_{\rm I}&=&\frac{\Delta\omega^{\rm 1PN}_{\rm tot}}{2\pi}+\frac{G^2m^2}{c^4 a_0^2(1-e_0^2)^3}\le[\frac34\le(86 +57 e_0^2 -13 e_0^4\ri)+3e_0\le(17+13e_0^2\ri)\cos f_0+\frac{15}2 e_0^2  \cos 2f_0\ri]\;,
\eqna
where $\Delta\omega^{\rm 1PN}_{\rm tot}$ is shown in \eqref{ior4}. The total precession $k_{\rm I}$ must be solely a function of the total orbital energy ${\cal E}$ of the particle and its angular momentum $h$ per unit mass which are equivalent to the mean values of the semi-major axis $\mathtt{a}$ of the particle's orbit and its eccentricity $\mathtt{e}$ respectively. However, the 2PN terms in the right hand side of \eqref{ior5} do not coincide with the formula \eqref{ior2}, and, hence, $k_{\rm I}=k_{\rm I}(\mathtt{a},\mathtt{e},f_0)=k_{\rm I}({\cal E},h,f_0)$ which means that it depends on three parameters instead of two. Therefore, $k_{\rm I}$ is not a radial integral of motion in contradiction to what the Hamilton-Jacobi method of integration of the particle orbit says -- see \eqref{6} and \eqref{7}. 
\end{itemize}

\section{Conclusion}\label{sec6}
The present paper provides a clear evidence of misconceptions and serious mathematical errors overwhelmingly present in the paper by \citet{iorio_2020}. The errors stem from the misuse of the method of successive iterations applied by \citet{iorio_2020} for solving the Gauss equations for osculating orbital elements of two-body system in the 2PN approximations. Numerical simulations conducted by \citet{iorio_2020} for testing his 2PN precession formula \eqref{ior5} are based on the same erroneous logic which was implemented for analytic calculations. For this reason, the numerical results of paper \citep{iorio_2020} merely repeat the analytic errors and, as such, are misleading and useless.   

The correct formula for the 2PN pericenter precession is either \eqref{10} and \eqref{11} or \eqref{24} depending on the choice of the constants of integration. In case of BepiColombo mission the correct formula for measuring the secular 2PN perihelion advance is given by equation \eqref{ior2}.
 
\begin{acknowledgement}
{\bf Acknowledgment}\newline
I am thankful to G. Sch\"afer, C.M. Will and R. Capuzzo Dolcetta for stimulating conversations. 
\end{acknowledgement}  

\bibliographystyle{unsrtnat}
\bibliography{2PN_pericenter_bib}

\end{document}